\documentclass[conference]{IEEEtran}
\IEEEoverridecommandlockouts
\usepackage{cite}
\usepackage{amsmath,amssymb,amsfonts}
\usepackage{algorithmic}
\usepackage{graphicx}
\usepackage{textcomp}
\usepackage{xcolor}
\def\BibTeX{{\rm B\kern-.05em{\sc i\kern-.025em b}\kern-.08em
    T\kern-.1667em\lower.7ex\hbox{E}\kern-.125emX}}
\begin{document}

\title{Identifying Key Expert Actors in Cybercrime Forums Based on their Technical Expertise\thanks{Study published in the 2024 APWG Symposium on Electronic Crime Research (eCrime) available at: www.doi.org/10.1109/eCrime66200.2024.00019}\\
}

\author{\IEEEauthorblockN{Estelle Ruellan}
\IEEEauthorblockA{\textit{Université de Montréal} \\
\textit{Flare}\\
Canada
}
\and
\IEEEauthorblockN{François Labrèche}
\IEEEauthorblockA{\textit{Secureworks} \\
Canada \\
}
\and
\IEEEauthorblockN{Masarah Paquet-Clouston}
\IEEEauthorblockA{\textit{Université de Montréal} \\
\textit{Complexity Science Hub}\\
Canada}
}

\maketitle

Study published in the 2024 APWG Symposium on Electronic Crime Research (eCrime) available at: www.doi.org/10.1109/eCrime66200.2024.00019. © 2025 IEEE. Personal use of this material is permitted. Permission from IEEE must be obtained for all other uses, in any current or future media, including reprinting/republishing this material for advertising or promotional purposes, creating new collective works, for resale or redistribution to servers or lists, or reuse of any copyrighted component of this work in other works.

\begin{abstract}
The advent of Big Data has made the collection and analysis of cyber threat intelligence challenging due to its volume, leading research to focus on identifying key threat actors; yet these studies have failed to consider the technical expertise of these actors. Expertise, especially towards specific attack patterns, is crucial for cybercrime intelligence, as it focuses on targeting actors with the knowledge and skills to attack enterprises. Using CVEs and CAPEC classifications to build a bimodal network, as well as community detection, k-means and a criminological framework, this study addresses the key hacker identification problem by identifying communities interested in specific attack patterns across cybercrime forums and their related key expert actors. The analyses reveal several key contributions. First, the community structure of the CAPEC-actor bimodal network shows that there exists groups of actors interested in similar attack patterns across cybercrime forums. Second, key actors identified in this study account for about 4\% of the study population. Third, about half of the study population are amateurs who show little technical expertise. Finally, key actors highlighted in this study represent a promising scarcity for resources allocation in cyber threat intelligence production. Further research should look into how they develop and use their technical expertise in cybercrime forums. 
\end{abstract}


\section{Introduction}
Cybercrime is an omnipresent threat in today’s digital age, causing significant financial losses and disrupting organizations worldwide~\cite{world_economic_forum_global_nodate}. Traditionally, criminal investigations have taken a reactive stance, focusing on specific incidents~\cite{eck_new_2019, samtani_using_2016}. However, the extensive damages caused by cybercrime stresses the need for a more proactive approach~\cite{geers_challenge_2010}.

Moreover, the exponential growth of data has made the production of cybercrime intelligence time-consuming and resources intensive. As a result, identifying relevant threat actors in cybercrime forums has become increasingly challenging~\cite{marin_mining_2018, huang_hackerrank_2021}. Such challenge is known as \textit{the key hacker identification problem}~\cite{marin_mining_2018}. 

This identification problem has drawn attention in both the computer science and criminology research fields ~\cite{holt_techcrafters_2008, benjamin_securing_2012, abbasi_descriptive_2014,zhang_classification_2015, fang_exploring_2016, samtani_using_2016, grisham_identifying_2017, marin_community_2018, marin_mining_2018, johnsen_identifying_2020, huang_hackerrank_2021}. However, when identifying key actors, no studies have focused on finding those with technical expertise, that is expertise towards cyber attack patterns. Expertise refers to ``the characteristics, skills and knowledge that distinguish experts from beginners and the less experienced” (p.3-4)"~\cite{ericsson_influence_2006}. Experts are therefore those with advanced knowledge on a topic compared to their peers. Plus, having expertise in a field has been found to be closely related to criminal success ~\cite{bartol_criminal_2014}. Hence, finding actors with technical expertise in attack patterns within cybercrime forums can focus intelligence resources  on those most likely to succeed in their criminal activities.

Using a framework from criminology~\cite{bouchard_professionals_2011}, as well as cluster analyses, this study addresses the key hacker identification problem by identifying communities interested in specific attack patterns across cybercrime forums and their related key actors. Key actors, in this study, are those who display high technical expertise in their communities.  

To do so, communities interested in similar attack patterns are identified by selecting posts in cybercrime forums mentioning Common Vulnerabilities and Exposures (CVEs)\footnote{https://cve.mitre.org/} and their corresponding Common Attack Pattern Enumeration and Classifications (CAPECs)\footnote{https://capec.mitre.org/about/index.html} from MITRE\footnote{https://www.mitre.org/}. Then, communities are formed by running the Leiden community-detection algorithm on a bimodal network linking actors with the CAPECs they mention in their posts. The meaning of communities is subsequently uncovered using content analysis. 

Then, key actors within these communities are identified based on the expertise they exhibit in each community. Such expertise is measured through two facets (as conceptualized in~\cite{bouchard_professionals_2011}): (1) actors' skill level and (2) commitment towards their respective community. A third element is added to the analysis: activity rate, to determine the extent to which key actors are active in cybercrime forums.  The framework also allows the categorization of the study population based on their technical expertise from professionals (high skill and high commitment: the key expert actors), pro-amateurs (high skill, little commitment), average career criminals (low skill, high commitment) and amateurs (low skill and low commitment). 

The study's key takeaways are:

\begin{itemize}
  \item The actor-CAPEC bimodal network displays a community structure that groups actors interested in similar attack patterns together. 
  % \item Attack patterns act as catalysts of cybercrime communities, transcending forum boundaries. 
   \item Actors with high skills and commitment in their community represent a tiny proportion (4\%) of the study population.
   \item About half of the study population are amateurs with regards to their technical expertise.

% In the case of cybercrime and even criminology, one’s expertise in a specific area should be nuanced by the diligence of one’s cybercriminal activities over time (activity rate)
\end{itemize}.

In sum, this study provides a method to find communities of interest in similar attack patterns based on CVE/CAPEC posts in cybercrime forums and find their key actors based on their technical expertise. The results show that key expert actors represent a promising scarcity for resources allocation in the production of cyber threat intelligence. Further studies should look into these actors to better understand their influence and behaviors in cybercrime forums. 

The rest of the paper is organized as follows: Section~\ref{sec:Literature} presents a literature review on the key hacker identification problem and its related field of research; Section~\ref{sec:Methodology} outlines the methods and data; Section~\ref{sec:Results} presents the results of the study; Section~\ref{sec:Discussion} provides a discussion; Section~\ref{sec:Limits} presents the limitations and future research; Section~\ref{sec:Conclusion} is the conclusion. 

\section{Literature Review}
\label{sec:Literature}
This section summarizes research on the key hacker identification problem, discusses the meaning of expertise and presents the framework.

\subsection{Identifying Key Actors in Cybercrime Forums}

Identifying a small number of relevant actors in cyber threat intelligence enables more efficient targeting of efforts and resources. Cybercrime communities consist of members with varying knowledge levels, and it is the more skilled and reputable members that are of primary interest~\cite{motoyama_analysis_2011, marin_mining_2018,bartol_criminal_2014}. These members represent only a small fraction of the cybercrime forum population, with the majority being unskilled or merely curious~\cite{marin_community_2018}. Three approaches have been developed to identify such key actors in cybercrime forums.

The first approach is based on social network analysis and focuses on identifying actors with significant centrality measures~\cite{decary-hetu_social_2012, samtani_using_2016, samtani_exploring_2017}. Actors are connected to each other through their social interactions using mono and bipartite networks and key members are uncovered with degree and betweenness centralities~\cite{decary-hetu_social_2012, samtani_using_2016, samtani_exploring_2017}. Using this approach, key actors are typically the most senior and active members~\cite{samtani_using_2016, samtani_exploring_2017}.

The second approach focuses on analyzing discussions (i.e., content analysis), generally establishing metrics to assess the activity and/or quality of interactions, and subsequently, identifying actors excelling in these metrics~\cite{holt_subcultural_2007, benjamin_securing_2012, zhang_classification_2015, fang_exploring_2016}. Key actors are identified based on their behaviors and how they are perceived in forums, considering, for example, the quality of the content they share, their reputation, their level of activity or their seniority~\cite{ benjamin_securing_2012, fang_exploring_2016}. Among the main findings, key actors are usually very active members~\cite{benjamin_securing_2012, abbasi_descriptive_2014, zhang_classification_2015} with the most seniority ~\cite{abbasi_descriptive_2014, zhang_classification_2015}. Their discussion is specialized in a particular~\cite{benjamin_securing_2012, abbasi_descriptive_2014, fang_exploring_2016} and often sophisticated topic ~\cite{abbasi_descriptive_2014, zhang_classification_2015}. They are the main knowledge-sharing actors within their community and do not hesitate to share cybercriminal assets in their messages ~\cite{benjamin_securing_2012, abbasi_descriptive_2014, zhang_classification_2015}.

The third research approach combines the previous two: social network and content analysis to identify key actors~\cite{abbasi_descriptive_2014, grisham_identifying_2017, marin_mining_2018 ,johnsen_identifying_2020, huang_hackerrank_2021}.  This hybrid approach suggests that key actors are also those with the most seniority and post the most~\cite{abbasi_descriptive_2014, grisham_identifying_2017,johnsen_identifying_2020, huang_hackerrank_2021}. They occupy a high rank or position within the forum~\cite{grisham_identifying_2017, johnsen_identifying_2020, huang_hackerrank_2021} and demonstrate great influence on their forum’s network, which is illustrated by high centrality measures~\cite{abbasi_descriptive_2014, grisham_identifying_2017, huang_hackerrank_2021}. The content they contribute is often high quality and reflects themes that are unique to each key actor~\cite{huang_hackerrank_2021}.

Hence, past research has developed different methods to pinpoint key actors within cybercrime forums. However, none of them have focused on identifying \textit{experts} in attack patterns \textit{across cybercrime forums}. Expertise in attack patterns is crucial for cybercrime intelligence, as it focuses on targeting actors with the knowledge and skills to attack enterprises.

\subsection{About Expertise}
Expertise can be defined as “the characteristics, skills and knowledge that distinguish experts from beginners and the less experienced” (p.3-4)~\cite{ericsson_influence_2006}. It refers to a wide range of cognitive knowledge and/or specialized skills in a specific domain~\cite{ericsson_influence_2006, van_gog_expertise_2012, nee_understanding_2015, bartol_criminal_2014}. To distinguish experts from others, the relative approach uses a continuum from novices to experts~\cite{hoffman_eliciting_1995, hoffman_how_1998, chi_two_nodate, nee_review_2015}. Reaching and maintaining the extreme end of the expertise continuum requires ongoing practice to keep one’s knowledge and skills in one’s chosen field up to date~\cite{ericsson_ericsson_nodate}.

For decades, expertise has been studied from multiple angles. Research in fields like law has examined expertise in professions such as chess players, pilots, and doctors~\cite{simon_skill_1988, schmidt_cognitive_1990, vicente_ecological_1998}. In the realm of criminal behavior, early studies focused on nonviolent offenses like burglary~\cite{wright_how_1988, wright_criminal_1995}, with subsequent research expanding to include expertise in areas such as violent crime~\cite{topalli_criminal_2005}, arson~\cite{butler_scripts_2015}, carjacking~\cite{topalli_it_2015}, and identity theft~\cite{vieraitis_little_2015}.

Moreover, expertise is linked to, and sometimes even necessary for criminal success~\cite{bartol_criminal_2014}. Individuals with criminal expertise in their field carry out their misdeeds in a more sophisticated manner and are more likely to be successful, making them more effective and dangerous~\cite{bartol_criminal_2014}. However, criminal success is evaluated reactively, relying on past actions to determine effectiveness. This reactive approach contrasts with proactive intelligence efforts aimed at prevention. In the absence of criminal success indicators, expertise, which can be assessed proactively, becomes essential for identifying relevant actors in cyber threat intelligence. 

In this study, experts are identified with the help of a framework~\cite{bouchard_professionals_2011}  developed in criminology and presented below. 

\subsection{Identifying Key Actors using Bouchard and Nguyen (2011)'s criminological framework}

Bouchard and Nguyen's ~\cite{bouchard_professionals_2011} framework revisit how to define professional criminals, which are, as stated by the authors: ``experts, the best in the field" (p.111). Drawing on research into professional crime ~\cite{hobbs_professional_1997, sutherland_professional_1937} and criminal success ~\cite{bouchard_professionals_2011}, the authors develop a classification using two criteria: skill level and commitment. Skill level refers directly to an individual’s expertise in their criminal field. The authors illustrate this concept using the analogy of professional athletes who possess specific skills and knowledge specific to their sport ~\cite{bouchard_professionals_2011}.  Commitment, on the other hand, implies that to become a professional, extensive involvement in the criminal activity is needed.

To distinguish experts from others, the authors develop a classification that includes four categories. The categories are defined based on skill level and commitment criteria, as presented in Table \ref{tab:bouchard_nguyen_framework}. Experts, dubbed professionals by the authors, are those who are highly skilled and committed. Average career criminals are committed but unskilled. Pro-amateurs are skilled but uncommitted and amateurs are unskilled and uncommitted; they are the novices. 

In summary, their framework categorizes individuals by a dual construction of expertise, considering both skill level and commitment. This dual construction makes their framework particularly relevant for identifying key actors in cybercrime forums based on their discussions on attack patterns. 

\begin{table}[ht]
    \centering
    \caption{Bouchard and Nguyen (2011, p. 111) Framework}
    \label{tab:bouchard_nguyen_framework}
    \begin{tabular}{|c|c|c|}
    \hline
      & High Commitment & Low Commitment \\ \hline
    High Skill Level & Professional [experts]& Pro-Amateur \\ \hline
    Low Skill Level & Average Career Criminal & Amateur \\ \hline
    \end{tabular}
\end{table}

\subsection{The Present Study: Identifying Key Actors Based on Their Expertise}
As discussed above, different approaches have been developed to identify key actors in cybercrime forums ~\cite{holt_techcrafters_2008, benjamin_securing_2012, abbasi_descriptive_2014,zhang_classification_2015, fang_exploring_2016, samtani_using_2016, grisham_identifying_2017, marin_community_2018, marin_mining_2018, johnsen_identifying_2020, huang_hackerrank_2021}. However, none of these studies have developed a method to identify key actors based on their expertise in attack patterns. Such expertise matters for cybercrime intelligence as it is required to attack enterprises. Expertise is also closely bound to criminal success ~\cite{bartol_criminal_2014} as it renders individuals more efficient, sophisticated, and successful in their misdeeds, making them more dangerous~\cite{bartol_criminal_2014}. 

Hence, drawing on Bouchard and Nguyen~\cite{bouchard_professionals_2011}'s two-faceted framework and previous literature, this study addresses the key hacker identification problem by, first, identifying communities interested in similar attack patterns across cybercrime forums and second, pinpointing their related key expert actors. To do so, two objectives have been developed. 

\begin{center} \textbf{Objective 1:} Identify communities interested in similar attack patterns across cybercrime forums. \end{center}

Attack patterns describe techniques, tactics and methods used to attack systems. Interest in similar attack patterns are identified by selecting posts in cybercrime forums mentioning CVEs and linking them with their corresponding CAPECs.

CVEs, or Common Vulnerability Exposures, are computer security vulnerabilities which, once exploited, allow entry into a system. Several organizations such as MITRE or NIST~\footnote{https://www.nist.gov/} keep a register of all CVEs known to date.  Each CVE has a unique standardized identifier (e.g. CVE-2019-12255). CAPEC, or Common Attack Pattern Enumeration and Classification, is a publicly available and community-driven catalog of known attack patterns. These attack patterns describe techniques, tactics and methods used by malicious actors to exploit weaknesses in systems. Each CAPEC represents a distinct method that attackers might employ to compromise systems. Each CAPEC can have many related CVEs.  In sum, mentions of CVE linked with their corresponding CAPEC represent a proxy for interests in attack patterns. 

Communities of interests in attack patterns are then uncovered by computing a bimodal network linking actors with their corresponding CAPEC and running the Leiden community-detection algorithm on the network. Then, the second objective is:

\begin{center} \textbf{Objective 2}: Detect key expert actors within these communities \end{center}

Key actors are identified based on the technical expertise they exhibit in each community. Such expertise is measured through two facets: actor's skill level and commitment, as defined by Bouchard and Nguyen’s~\cite{bouchard_professionals_2011} framework. For the analysis, a third element is added: activity rate, which is an indicator of one's involvement in cybercrime forums. It is relevant considering the online nature of the interactions which means that an actor's activity on a topic can be quantified. Moreover, the literature highlights that 
key actors are among the most active of their community~\cite{benjamin_securing_2012, abbasi_descriptive_2014, zhang_classification_2015, samtani_exploring_2017, samtani_using_2016, grisham_identifying_2017,johnsen_identifying_2020, huang_hackerrank_2021}. 

The distribution of technical expertise across the study population is also presented based on the four categories presented in Bouchard and Nguyen’s~\cite{bouchard_professionals_2011} framework, from amateurs [novices] to professionals [experts].

\section{Methodology}
\label{sec:Methodology}
\subsection{Data Collection}

The data was collected using the Flare~\footnote{https://flare.io/} search engine API. Flare systems is an information technology (IT) security company that maintains a cyber threat intelligence platform by monitoring various online spaces. It has been used in previous literature~\cite{paquet-clouston_role_2021, paquet-clouston_entanglement_2022}.

To build a network of actors and their interest in attack patterns, all posts 1) coming from cybercrime forums and 2) mentioning a CVE were fetched. Specifically, all posts that mentioned “CVE-2023”, “CVE-2022”, “CVE-2021”, “CVE-2020” and “CVE-2019”. As actors in cybercrime forums tend to not be active for a long time~\cite{hughes_digital_2023}, the search was limited to CVE published within the past five years. Also, note that although the search was restricted to recent CVEs, the dataset also includes older CVE references found in posts when actors mentioned multiple CVEs (e.g. CVE-2022-22965 and CVE-2017-9798). Relevant information for each post was then extracted, including the actor, forum, timestamp, content, and the CVE mentions.

\subsection{The Dataset}
The data collected included 11,558 posts made by 4,441 different actors on 124 cybercrime forums. The posts dated from January 8, 2015, to July 31, 2023, and mentioned 6,232 different CVEs.

\begin{figure}[ht]
    \centering
    \includegraphics[width=0.5\textwidth]{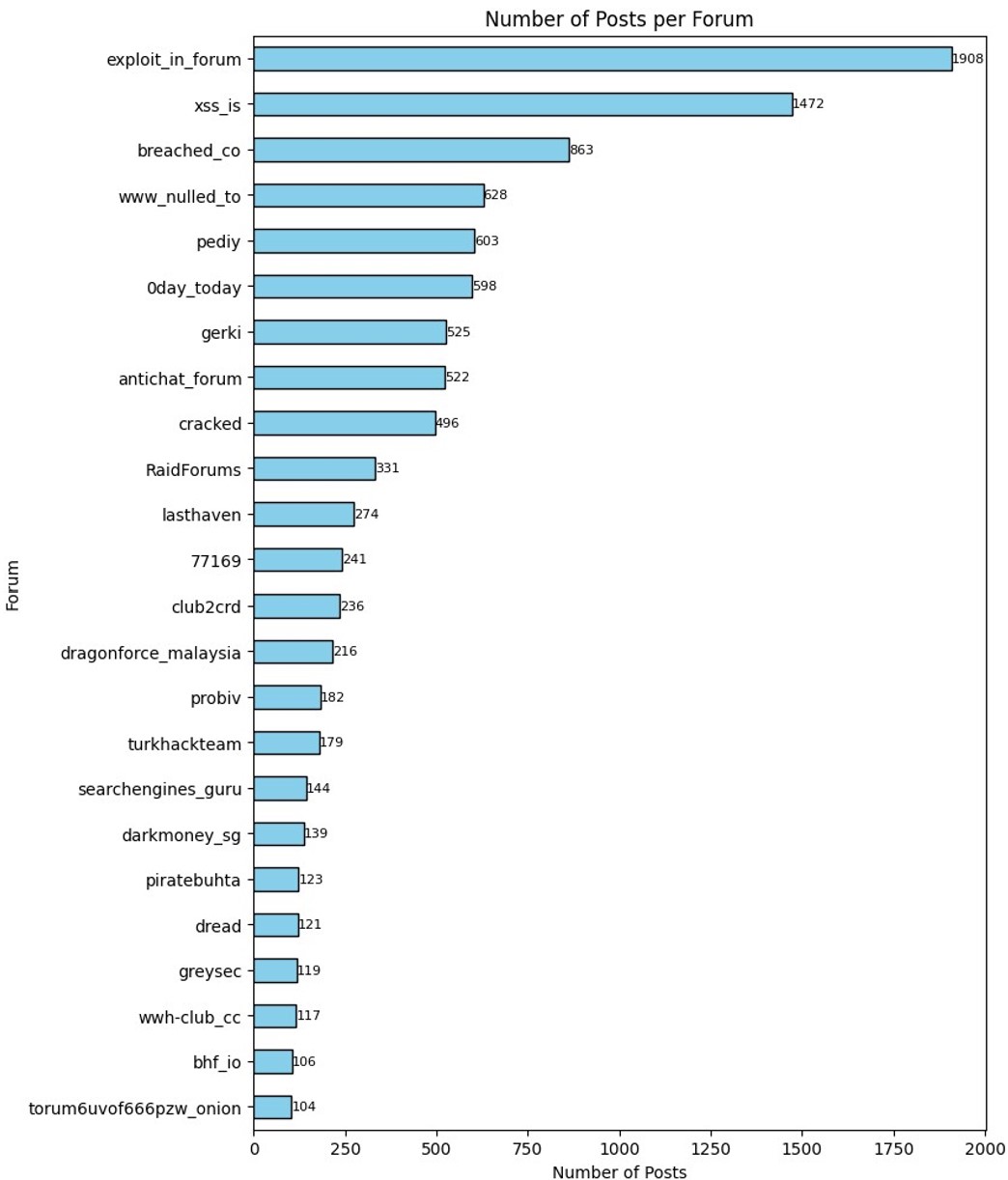}
    \caption{Number of Posts per Forum}
    \label{fig:Posts per Forum}
\end{figure}

Figure \ref{fig:Posts per Forum} displays the top 25 forums with the most posts. The forum with the most posts was exploit\_in with 1,908 posts (16.51\% of total posts), followed by xss\_is with a little less than 13\% of all posts. As shown in the figure, the distribution of posts across forums was uneven, since 99 forums had fewer than 100 posts and 43 of them had fewer than 10 posts. 

To achieve the first objective, we built an actor-CAPEC bimodal social network, as presented below.

\subsection{Bimodal Social Network}
Bimodal networks or two-mode networks are networks in which nodes can be categorized into two distinct groups or modes whereas a one-mode network only consist of one group of nodes~\cite{wasserman_social_1994}. A mode represents the nature of the node, in the case of this study, CAPEC vs. actor.

To do so, we used the CVEs mentioned by actors. Each actor was assigned the CAPECs corresponding to the CVEs they mentioned: if actor A mentioned CVE-2022-45451, then actor A was assigned the corresponding CAPEC of CVE-2022-45451, which is CAPEC 233. Some actors mentioned CVEs that did not have CAPECs and were therefore removed from the network. This reduced the number of actors from 4,441 to 3,308. Then, an unweighted bimodal network was built based on actor-CAPEC interactions. The mapping of CAPEC to CVE was done through their matching common weakness enumeration (CWE) identifiers. CWEs are a publicly available standardized naming of software weaknesses~\footnote{https://cwe.mitre.org/}.

\subsubsection{The 500-in-Degree Filter}
Some CAPECs were present in a large number of attacks. However, we sought to identify precise attacks used by subgroups of actors, and more general methods shared by a large number of actors were not beneficial to this aim. Hence, if a CAPEC was shared by a large number of actors, its relevance was minimal. CAPECs shared by a large number of actors were thus removed.

To do so, we filtered CAPEC mentioned by more than 500 actors, i.e., mentioned by more than 15\% of total actors in the network. The choice of a 500-in-degree threshold was informed by the observation of an apparent ‘elbow’ in the data, where the in-degree distribution exhibited a notable change. 

Fifty-seven CAPECs exceeded the 500-in-degree threshold and were removed. Of those 57 CAPECs, 55 (97\%) of them were general attack patterns and/or very popular ones (used by most actors), such as “Using Slash in Using Slashes” (CAPEC 79, 64, 78, 76) or “XML Injection” (CAPEC 250). Only 2 were outsiders: CAPEC 230 (Serialized payload data) and 231 (Oversized serialized payload data) . However, they both had 670 different actors mentioning them, i.e, 28.74\% of all actors; making them very popular. They were therefore removed.

This 500-degree filter eliminated 987 actors; their CAPECs having been eliminated from the analysis. Since these actors no longer had any CAPECs mentions, they were also removed from the database reducing the dataset to 2,321 actors and 263 CAPECs.

\subsection{Final Dataset}
The bimodal actor-CAPECs network was built with this filtered dataset and in-degree and out-degree centralities were computed. Within this network, actors mentioning CAPEC X were linked to the latter. The network allowed the visualization of relationships between actors and CAPECs, as well as those between actors through the exploitation of similar CAPECs. These relationships were the foundations of the identification of communities of interest towards attack patterns. 

The network included 263 CAPECs and 2,321 actors from 116 different forums. Table  \ref{tab:actors_overview} presents statistics on the network and the dataset. The one-timer variable highlight actors with only one post.

\begin{table}[ht]
    \centering
    \caption{Actors Overview}
    \label{tab:actors_overview}
    \begin{tabular}{|p{2cm}|c|c|c|c|c|c|}
    \hline
    & Count & Mean & Std & Min & Median & Max \\ \hline
    Actor Out Degree& 2321 & 13.40 & 23.46 & 1 & 3 & 187 \\ \hline
    Nb Posts CVE /CAPEC& 2321 & 3.51 & 13.23 & 1 & 1 & 375 \\ \hline
    One Timer & 2321 & 0.56 & X & 0 & X & 1 \\ \hline
    Nb of  Posts CVE/CAPEC without One Timers& 1006 & 6.80 & 19.63 & 2 & 3 & 375 \\ \hline
    \end{tabular}
\end{table}

Actors were linked with, on average, 13 different CAPECs. At least 50\% of actors had posted only one post with a CVE and a corresponding CAPEC,  and the maximum is 375.  Overall, 56.70\% of actors posted only once and among those who posted multiple times, the average number of CVE/CAPEC posts was six.

\subsection{Identifying 
Communities Interested in Similar Attack Patterns}

To achieve the first objective, we applied the Leiden community detection algorithm to the actor-CAPEC network. The Leiden community detection algorithm~\cite{traag_louvain_2019} yields high-quality partitions across various network scales without requiring manual parameterization. Its selection was driven by its prevalence in criminological research alongside the Louvain algorithm~\cite{calderoni_communities_2017,schaefer_friends_2017, paquet-clouston_robust_2023}. The quality of the community structure is measured with the modularity measure~\cite{newman_finding_2004}, which assesses the strength of community structures by comparing intra-community and inter-community connections.

\subsubsection{Content Analysis: the Shared Interest Behind Each Community}
To make sense of the communities identified by the Leiden algorithm, we used content analysis. Content analysis is a widely used research method and consists of systematically analyzing the content of textual, visual, or audio information to identify, code and categorize recurring patterns or elements~\cite{krippendorff_content_2018, hsieh_three_2005}. 

To assess the coherence of the Leiden communities, we analyzed their CAPECs to determine if they represented meaningful associations or arbitrary groupings. This involved examining each CAPEC's description, relationships, domain, and attack mechanism. Through this analysis, we identified distinct themes and categories within each community and assigned nomenclature based on their specific interests in certain attack patterns.

\subsection{Detect Key Actors Based on Their Technical Expertise}
To achieve the second objective, key actors were identified based on the expertise they exhibit in each community. Such expertise was measured through two facets: actor's skill level and commitment. A third variable: activity rate was calculated to measure actors' level of involvement in cybercrime forums. 

\subsubsection{Skill Level}
The skill level was measured based on the CAPECs associated with each actor in the bimodal network. Using MITRE's `Skill Level Required' metric, which categorizes `Skills Level Required' to execute the attack as `Low', `Medium', or `High', we assigned each CAPEC its highest skill level scenario. This approach prevented underestimating actors' skills and ensured that all potentially important actors were included.

Each actor thus had a list of skill level values, based on their associated CAPECs in the network. In short, if an actor mentioned four CVEs, and those CVEs were linked to six CAPECs, this actor's list would consist of six skill level values ranging from `Low' to `High'.

The overall distribution of skill level values among the actors' list is available in Table \ref{tab:skill_level_distribution} below. Additionally, Table \ref{tab:skill_level_values} provides further statistics on the distribution.

\begin{table}[ht]
    \centering
    \caption{Overall Distribution of Skill Level Values}
    \label{tab:skill_level_distribution}
    \begin{tabular}{|l|c|p{3cm}|}
    \hline
    Skill Level Value & Nb CAPECs & \% of Skill Level Values Among All Values in Actors' List \\ \hline
    Low & 118 (44.87\%) & 57.71\% \\ \hline
    Medium & 66 (25.09\%) & 24.14\% \\ \hline
    High & 79 (30.04\%) & 18.14\% \\ \hline
    \end{tabular}
\end{table}

\begin{table}[ht]
    \centering
    \caption{Skill Level Values Proportion Statistics}
    \label{tab:skill_level_values}
    \begin{tabular}{|p{1.4cm}|p{1.8cm}|c|c|c|}
    \hline
    Skill Level Value & Average Proportion in actor’s list & Median & 75th Percentile & Std \\ \hline
    High & 29.07\% & 23.08\% & 50.00\% & 30.76\% \\ \hline
    Medium & 36.12\% & 30.77\% & 50.00\% & 32.41\% \\ \hline
    Low & 33.74\% & 33.33\% & 66.66\% & 31.72\% \\ \hline
    \end{tabular}
\end{table}

Each actor’s list was transformed into a numerical list where `Low’=1, `Medium’=2 and `High’=3. %This way, an actor’s list went from [‘Low’, ‘Medium’, High’] to [1, 2, 3].

To establish a representative skill level, we used the 70th percentile value from each actor's list of skill levels. This choice reflects the conceptual idea that an actor with a significant proportion of `High’ values was more technically proficient than one with only `Medium’ and `Low’ values. It also ensured that actors with a significant proportion of `High' values were perceived as more technically proficient. Compared to weighted mean and median alternatives, the 70th percentile better reflected actors' skills given the distribution imbalance, where `High' values were less frequent. This aligned with the study's goal of identifying key actors with elevated skill levels, ensuring that only those with over 30\% `High' values were classified as highly skilled.

\subsubsection{Commitment Level}
Commitment was measured through actors' focus within their communities of interests (CoI). It was measured by the percentage of an actor's posts referencing CAPEC entries within their CoI relative to their total posts. In short, a post was categorized as `in-interest' if the majority ($x >= 50\%$) of its referenced CAPECs belonged to the actor's CoI. Then, commitment level was quantified by the proportion of `in-interest' posts relative to an actor's total posts. 

\begin{table*}[ht]
\centering
\caption{Descriptive Statistics of the Number of Specialized Posts per Actors without One Timers}
\begin{tabular}{|c|c|c|c|c|c|c|c|}
\hline
\textbf{Variable} & \textbf{Count} & \textbf{Mean} & \textbf{Std} & \textbf{Median} & \textbf{60th Percentile} & \textbf{75th Percentile} & \textbf{Max} \\ \hline
Nb specialized posts without one timers & 1006 & 6.80 & 19.63 & 3 & 4 & 5 & 375 \\ \hline
\end{tabular}
\label{tab:descriptive_stats_posts}
\end{table*}

As shown in Table~\ref{tab:descriptive_stats_posts}, 50\% of our actors have three or fewer specialized posts. Given this distribution, we removed all actors with fewer than four posts to ensure meaningful commitment levels. Filtering out actors with fewer than four specialized allowed to keep a substantial proportion of the population while getting commitment levels that are easier to work with. This reduced the final dataset to 359 actors.

\subsubsection{Activity Rate}

Activity rate is a third element added to Bouchard~\cite{bouchard_professionals_2011}’s framework, as mentioned above. It was measured by dividing the number of posts with a CVE and a corresponding CAPEC divided by the actor's total activity time in number of days. Hence, an actor with 10 posts spanning from March 20, 2021, to December 20, 2021 (totaling 275 days), would have an activity rate of 0.036 (10 posts / 275 days). This third element allowed understanding the extent to which an actor was active in cybercrime forums. 

% Activity rate, calculated as the ratio of specialized posts to the total activity time in days, measures an actor's posting activity relative to their seniority. For example, 

\subsubsection{Sample for the Identification of Key Actors}
Table \ref{tab:descriptive_statistics} presents descriptive statistics on the skill level, commitment and activity rate of actors. The sample for the identification of key actors consisted of 359 actors. The average actor had 36.68\% of posts committed to their CoI and had a skill level of 2.19 (`Medium'). The average activity rate was 0.72 (std=1.90).

\begin{table*}[ht]
    \centering
    \caption{Descriptive Statistics of Sample}
    \label{tab:descriptive_statistics}
    \begin{tabular}{|l|c|c|c|c|c|c|}
    \hline
    & Mean & Std & Min & Median & 75th Percentile & Max \\ \hline
    Length of Skill Level values list & 99.42 & 255.76 & 4 & 25 & 85 & 3449 \\ \hline
    Skill Level 70th percentile value & 2.19 & 0.64 & 1 & 2 & 3 & 3 \\ \hline
    Nb of posts (CVE with CAPEC)& 14.55 & 31.37 & 4 & 6 & 10 & 375 \\ \hline
    \% commitment & 36.68 & 29.61 & 0 & 25 & 50 & 100 \\ \hline
    Activity time (days) & 449.07 & 545.02 & 1 & 227.00 & 690.00 & 2669.00 \\ \hline
    Activity rate & 0.72 & 1.90 & 0.002 & 0.04 & 0.20 & 14.00 \\ \hline
    \end{tabular}
\end{table*}

\subsubsection{Finding Key Actors}
% Actors fitting the 'Professional' class (i.e. showcasing the highest technical expertise, combining high skill level, commitment to their Community of Interest (CoI)) and striking the right balance between seniority and diligence are thus considered key actors in our framework. 

To identify key actors, i.e., actors who score high in skill level, commitment and activity rate, we used the K-means clustering algorithm~\cite{jain_data_2010}. The K-means algorithm groups data points into a partition of a predefined number of clusters to discover patterns and structure in the data. This clustering is based on iteratively minimizing the Euclidian distance between each data point and the centroid of its assigned cluster.

\section{Ethical Considerations}
%  at the University of Montréal (project 2023-4678)
The study was approved by the university's ethics committee under minimal risks [study number 2023-4678]. It required asking for a waiver of consent in line with Article 5.5A of the Canadian Tri-Council Policy Statement on research ethics. To ensure user confidentiality and privacy, actors' pseudonyms are not displayed throughout the text.
% . 

\section{Results}
\label{sec:Results}
This section presents the key findings of this research. It starts with a presentation of the bimodal network actor-CAPEC. Then, the communities of interest are presented, followed by the distribution of key expert actors within these communities. 

\subsection{The Bimodal Actor-CAPEC Network}
The actor-CAPEC bimodal network has 2,584 nodes (2,321 actors and 263 CAPECs), and 31,093 edges. The network has a mean bilateral degree (in and out degrees combined) of 24 and a density of 0.009, meaning that less than 1\% (0.9\%) of possible connections between nodes are in the network. On average, actors are connected to 13 different CAPECs while CAPECs have links with 118 actors, as shown in Table \ref{tab:actor_capec_network}. 

\begin{table}[ht]
    \centering
    \caption{Actor-CAPEC Network Characteristics}
    \label{tab:actor_capec_network}
    \begin{tabular}{|c|c|c|c|}
    \hline
    & Count & Mean Degree & Std Degree \\ \hline
    Actors & 2321 & 13.40 & 23.46 \\ \hline
    CAPECs & 263 & 118.22 & 119.40 \\ \hline
    \end{tabular}
\end{table}

\subsection{Communities of Interest in Attack Patterns}
\begin{figure}[ht]
    \centering
    \includegraphics[width=0.5\textwidth]{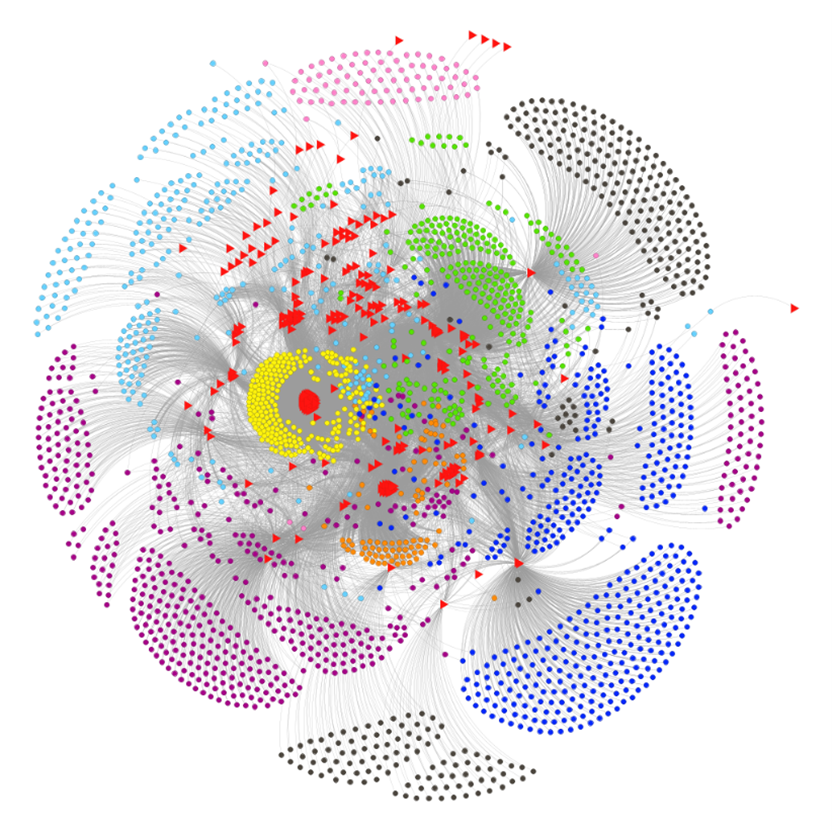}
    \caption{Bimodal actor-CAPEC Network Colored according to Communities of Interests. Note. The representation of the graph uses the Fruchterman Reingold projection with the following settings in Gephi: zone=10000; Gravity=7.0; Speed=5.0.}
    \label{fig:Bimodal Colored}
\end{figure}

\begin{figure}[ht]
    \centering
    \includegraphics[width=0.5\textwidth]{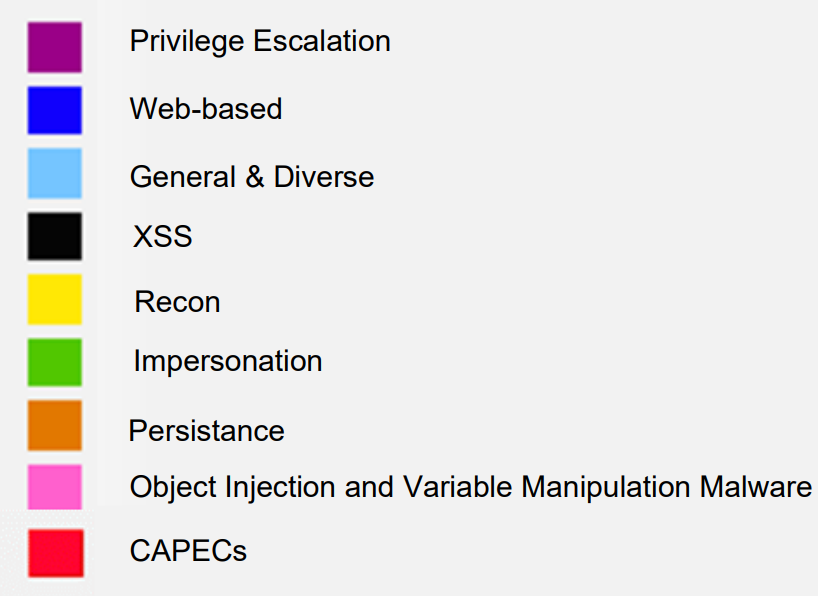}
    \caption{Bimodal actor-CAPEC Network Colored Legend}
    \label{fig:Bimodal Colored Legend}
\end{figure}

Through iterative application of the Leiden algorithm, we identified eight distinct communities within the network. The final partition achieved a modularity score of 0.473, the highest among the iterations, exceeding the well-established threshold of 0.3 and indicating a substantial level of cohesion within the identified communities~\cite{newman_finding_2004}. Figure \ref{fig:Bimodal Colored} depicts the network with the actors colored according to their respective communities. CAPECs are in red for clarity. The color legend for Figure \ref{fig:Bimodal Colored} is shown in Figure \ref{fig:Bimodal Colored Legend}.

To provide a better overview of the communities of interest, the communities’ characteristics are presented in Table \ref{tab:coi_overview}.
The average out-degree represents the number of unique CAPECs an actor is linked with. Then, each community is interpreted, and their respective characteristics are presented below.

\begin{table*}[ht]
    \centering
    \caption{Communities of Interest (CoI) Overview}
    \label{tab:coi_overview}
    \begin{tabular}{|p{1.2cm}|p{1.6cm}|c|c|c|c|p{2.0cm}|p{2.0cm}|p{2.0cm}|c|}
    \hline
    Community & Community of Interest & Nodes & CAPEC & actors & \% one timers & Mean out-degree per actor & Std (out-degree) & Mean nb of specialized posts & Std (posts) \\ \hline
    0 & PrivEsc & 544 & 19 & 525 & 65.14 & 4 & 7.11 & 2 & 4.76 \\ \hline
    1 & Web-based & 497 & 26 & 471 & 71.97 & 5 & 12.98 & 3 & 18.33 \\ \hline
    2 & General & 431 & 103 & 328 & 56.10 & 14 & 33.15 & 7 & 24.89 \\ \hline
    3 & XSS & 319 & 10 & 309 & 71.52 & 2 & 1.18 & 1 & 1.46 \\ \hline
    4 & Recon & 298 & 55 & 243 & 51.44 & 61 & 9.04 & 3 & 6.99 \\ \hline
    5 & Impersonation & 296 & 25 & 271 & 54.61 & 12 & 7.88 & 3 & 5.49 \\ \hline
    6 & Persistence & 116 & 22 & 94 & 41.49 & 26 & 25.76 & 5 & 7.96 \\ \hline
    7 & OIVMM & 83 & 3 & 80 & 85.00 & 1 & 0.31 & 1 & 1.62 \\ \hline
    \end{tabular}
\end{table*}

\textbf{Community 0: PrivEsc} The PrivEsc community is related to the privilege escalation attack pattern. Privilege escalation is when an attacker gains more access or control over a system than they should have. The PrivEsc community is the most populated with 525 actors, i.e., 22.62\% of all actors, but has the third least number of CAPECs (19). In this community, 65.14\% of actors are one timers. 

\textbf{Community 1: Web-Based} The Web-Based community is focused on web-based attacks. It counts 471 members, and 26 CAPECs, making it the second-largest community in our network. However, more than 70\% of its population is a one timer (71.97\%). 

\textbf{Community 3: XSS} The XSS community specializes in XSS attacks (cross-site scripting). XSS is when attackers inject harmful JavaScript code into a website or application that is then seen by other users. With a percentage of 71.52\% of one timers among its 309 actors, the XSS community holds third place in one timer percentage. The XSS community counts 10 CAPECs, which is the second-lowest number of CAPECs.

\textbf{Community 4: Recon} The Recon community only contains CAPECs about reconnaissance and scanning.  Recon and scanning involve gathering information about a target system to find weaknesses. It has the second higher number of CAPECs with 55. The percentage of one timers is also among the lowest with 51.44\%. The Recon members sit at the top for average number of 76 CAPECs: they have a link with 61 CAPECs, with an average number of posts of three. 

\textbf{Community 5: Impersonation} The Impersonation community includes impersonation attacks, combining CAPECs about authentication bypassing and spoofing. It is populated by 271 actors and 25 CAPECs. In this community, just over half (54.61\%) of actors are one timers. On average, members have links with 12 CAPECs and published three  posts.

\textbf{Community 6: Persistence} The persistence community is focused on persistence techniques related to either an attack or a type of malware. It is when attackers solidify the attack’s foothold on the system by writing it onto the disk to make sure their presence in a system lasts for a long time, even after the initial attack or a restart of the machine. The persistence community has the lowest percentage of one timers with 41.49\%. It counts 22 CAPECs. However, the members of this community have the second-highest average number of CAPEC they are linked with, with 26 CAPECs, and posted, on average, five times.

\textbf{Community 7: OIVMM} is related to object injection and variable manipulation malware (OIVMM) attacks. The OIVMM can lead to unauthorized access, information theft or control over the digital system. The OIVMM community is the smallest of our network. With only 80 actors and three CAPECs, this community holds the first place in terms of percentage of one timers with 85\% of its members having posted only once. 

And lastly, \textbf{Community 2} is a diverse community that is not linked to a specific attack pattern. This community contains a myriad of different CAPECs that are not specific to certain types of targets or attack patterns. With more than 100 CAPECs, this community seems to be the home of the only diverse/versatile community. Having 103 CAPECs places this community at the top for number of CAPECs. However, it places third for the number of actors with 328. This community has 56.10\% of one timers. 

Then, having established the distinct communities of interest based on shared CVE/CAPEC discussions, objective 2 aims at identifying key expert actors within them. 

\subsection{Unveiling the Spectrum Key Expert Actors}
To identify key expert actors, we used the K-means clustering algorithm on the three variables that shape the expertise concept: skill level, commitment and activity rate. 

\begin{figure}[ht]
    \centering
    \includegraphics[width=0.5\textwidth]{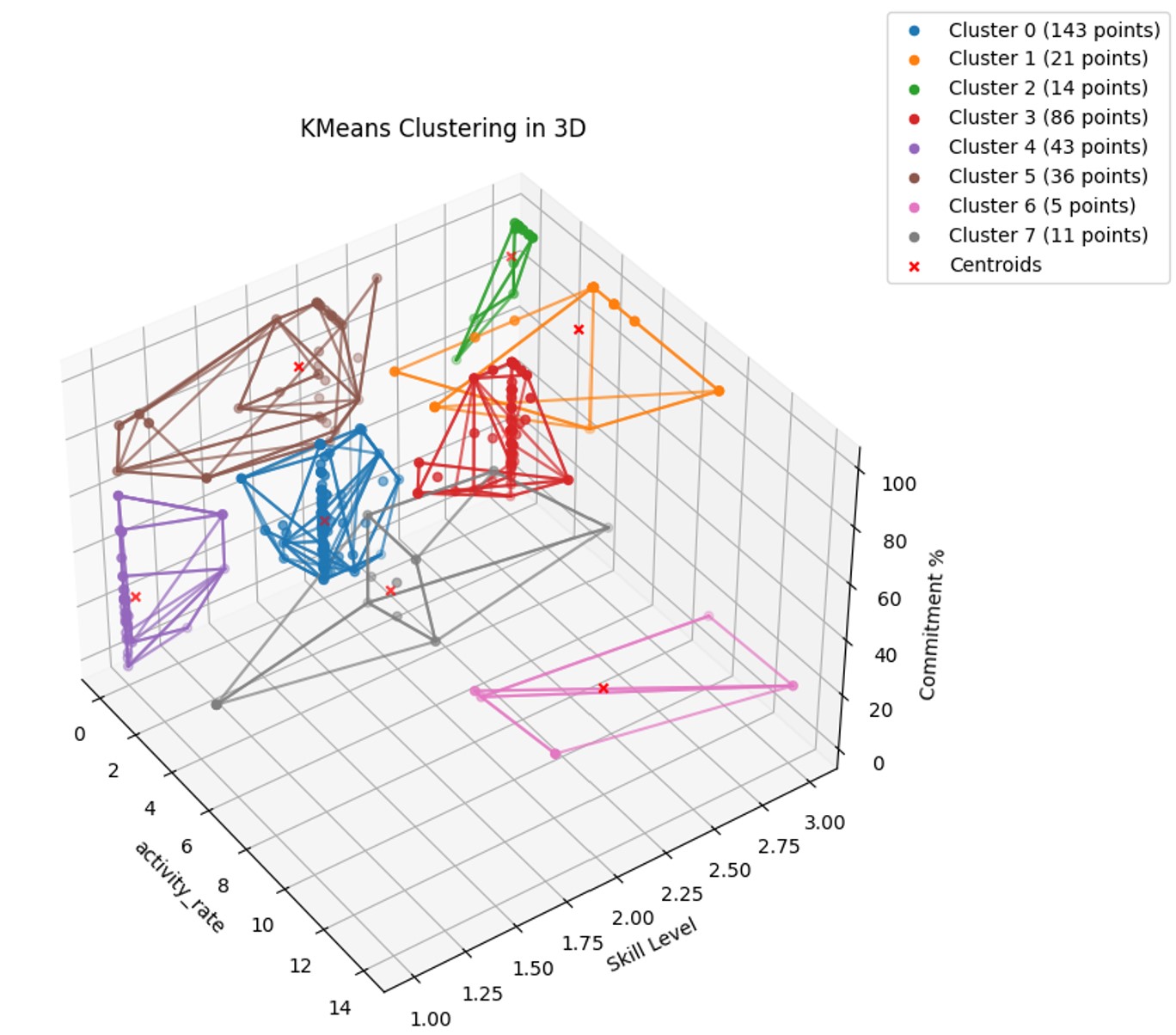}
    \caption{Partition of the model with k=8 clusters}
    \label{fig:K-means Partition}
\end{figure}

Our analysis identified an optimal model, one with eight clusters and a silhouette score of 0.569. Its partition is shown in Figure \ref{fig:K-means Partition}, with each cluster having its assigned color. Each cluster is plotted on a 3-dimensional space, with activity rate on the x-axis, skill level on the y-axis and commitment percentage on the z-axis.

We categorized clusters by  expertise levels based on their centroids’ score on skill level, commitment and activity rate. We use the terminology developed in Bouchard and Nguyen's~\cite{bouchard_professionals_2011}’s framework to name the clusters, thus facilitating their interpretation. Table \ref{tab:clusters_overview} provides an overview of the clusters.

\begin{table*}[ht]
    \centering
    \caption{Overview of Clusters}
    \label{tab:clusters_overview}
    \begin{tabular}{|c|p{4.0cm}|p{6.0cm}|c|c|}
    \hline
    Cluster & Bouchard \& Nguyen framework & Centroid [Skill; Commitment; Activity] & Nb of actors & \% of sample population \\ \hline
    0 & Amateurs & [2.00; 22.47; 0.11] [Mid; Low; Discrete] & 143 & 39.83 \\ \hline
    1 & Pro-Amateurs& [2.81; 97.62; 5.14] [High; High; Short-lived]& 21 & 5.85 \\ \hline
    2 & Professionals & [2.96; 90.37; 0.28] [High; High; Active] & 14 & 3.90 \\ \hline
    3 & Pro-Amateurs & [2.96; 25.32; 0.12] [High; Low; Discrete] & 86 & 23.96 \\ \hline
    4 & Amateurs & [1.05; 24.32; 0.05] [Low; Low; Discrete] & 43 & 11.98 \\ \hline
    5 & Average Career Criminals & [1.86; 84.81; 0.50] [Low; High; Active] & 36 & 10.02 \\ \hline
    6 & Pro-Amateurs & [2.38; 18.46; 10.67] [Mid; Low; Hyperactive] & 5 & 1.39 \\ \hline
    7 & Amateurs & [1.95; 24.51; 4.14] [Mid; Low; Hyperactive] & 11 & 3.06 \\ \hline
    \end{tabular}
\end{table*}

\subsubsection{Professionals: the Experts}
According to Bouchard and Nguyen's framework~\cite{bouchard_professionals_2011}, professionals are those with high skill level and commitment. They are ``the experts, the best in their field" (p.111). 

One cluster exhibits high scores in these two facets: cluster 2, as shown in Table \ref{tab:clusters_overview}. Cluster 2 gathers 14 individuals with high skills and high commitment, as well as low activity rate, as shown with its centroid [2.96; 90.37; 0.28]. With an average skill level of 2.96 and a commitment level of 90.37\%, these actors exhibit top-tier skills and a strong dedication to their community of interest. Their activity rate of 0.28 is low. The reason behind this is that they are older or even senior members, with a longer period of activity, making their activity rate plummet. With an average period of activity of 159 days and average posting rate of one post every three to four days, they strike the right balance to be considered key actors: they exhibit the highest level of expertise in our framework and are among the most senior members of their community. They represent 3.90\% (14 out of 359) of the sample.

\subsubsection{Pro-Amateurs}
Pro-Amateurs are actors scoring high on the skill level but relatively low on the commitment scale. Pro-Amateurs represent the second level of technical expertise, just below Professionals. Two clusters follow these criteria:  Cluster 3 and Cluster 6. We also categorized Cluster 1 as pro-amateurs, due to the nuance interpretation on commitment that the activity rate variable showed, as explained below. Hence, Pro-Amateurs account for 31.20\% of the study population.

Cluster 3 is characterized by a centroid with the highest skill level (2.96) but a commitment percentage barely above 25\% and one of the lowest activity rates (0.12). Hence, these actors have top-tier skills with a tendency to explore various attack patterns rather than focusing on one. Their low activity rate is also due to their long period of activity; on average they were active for 488 days with some having a track record of more than 2,500 days. However, despite their seniority, these actors don’t contribute frequently. With an average activity rate of 0.12, these actors tend to share specialized content intermittently. They represent 23.96\% of the sample. 

Cluster 6 gathers a more short-lived hyperactive population. The population of this cluster has a mid to top-tier skill level (2.38), meaning they oscillate between medium and high skilled CAPECs, and shows the lowest commitment percentage (18.46\%) of all clusters. Moreover, they have the highest activity rate by far (10.67), as they were active only for a day, meaning they post 10 times a day. This cluster consists of only 5 actors (i.e. 1.39\% of our sample).

Cluster 1 gathers 21 individuals with high skills, commitment and activity rate, as shown with its centroid [2.81; 97.62; 5.14]. With an activity rate of 5.14 on average, this population posts 5 times a day. The reason behind such a high activity rate is their limited period of activity. All actors in this population have only been active for a day in our sample. With an average of five CVE/CAPEC posts a day, despite being short-lived, these actors should be monitored as they exhibit high skills and high commitment. They represent pro-amateurs and account for 5.85\% of the sample.

\subsubsection{Average Career Criminals}
Average Career Criminals are actors scoring low on skill level but high in terms of commitment. A single cluster fits this class: Cluster 5. Its centroid indicates a low to mid-tier skill level (1.86), a high commitment (84.81\%) as well as a relatively high activity rate (0.5). Despite their skill level neighboring the mid-tier, these actors exhibit a high commitment to their attack pattern of interest, checking the boxes to be considered `Average Career Criminal' according to Bouchard and Nguyen’s classification. They constitute just over 10\% of our sample (10.02\%).

\subsubsection{Amateurs}
Amateurs is the last class of our framework. Amateurs are those who score low on both scales. The amateur population is the largest of our sample, with more than half (54.87\%) of our sample. It is scattered in three clusters: Cluster 0, Cluster 4 and Cluster 7.

Cluster 0's centroid presents a skill level of 2, a commitment of 22\% and an activity rate just above 0.10. These actors have mid-tier skills and can’t seem to settle for a single attack pattern of interest, hence their low commitment rate. Finally, they don’t offer a very active contribution and account for a little less than 40\% of our sample (143/359 = 39.83\%).

Cluster 4 represents the lowest skilled discrete amateurs. Characterized by a centroid with the lowest skill level (1.05) and activity rate (0.05) and a commitment percentage of 24.32\%, these actors possess the lowest skills and display a curiosity to explore various attack patterns. This population is also the least active of our clusters and gathers 11.98\% of the sample.

Cluster 7 represents the hyperactive amateur population with a centroid of medium skill level (1.95) as well as a low commitment (24.51\%). Nevertheless, cluster 7 exhibits a high activity rate with an average of 4.14. Hyperactive amateurs constitute 3.06\% of all actors.

\section{Discussion}
\label{sec:Discussion} 

The study yields three key findings: 1) The actor-CAPEC bimodal network displays a community structure that groups actors interested in similar attack patterns together. 2) Actors with high skills and commitment in their community represent a tiny proportion (4\%) of the study population. 3) About half of the study population are amateurs with regard to their technical expertise. These key takeaways are discussed below. 

\setcounter{subsubsection}{0} 
\subsubsection{The Community Structure Behind the Actor-CAPEC Bimodal Network }
First, it is interesting to note that groups of actors interested in similar attack patterns were uncovered with the analysis. Indeed, the Leiden community detection algorithm, while based on network structure and mathematical criteria~\cite{traag_louvain_2019, anuar_comparison_2021}, does not ensure meaningful community relevance to the research question at hand. Yet, we mapped the CAPEC-actor network using CVE mentions and their related CAPECs, acknowledging the potential for information loss in the process. Despite this, our content analysis showed that the communities identified by the Leiden algorithm reflected shared interests in specific attack patterns. Attack patterns behind the uncovered communities included PrivEsc, Web-Based, XSS, Recon, Impersonation, Persistence, Recon and OIVMM. Hence, by extrapolating from CVE posts and their related CAPEC, we found that actors do tend to post content about similar attack patterns, despite potential CVE-CAPEC association inaccuracies.  

This reveals a new dimension of cyber threat intelligence: the need to monitor communities interested in similar attack patterns. Further research should investigate how these communities are distributed in cybercrime forums and how they help the development of attack pattern expertise among its members.

\subsubsection{Unveiling Key Expert Actors for Targeted Intelligence}
Key actors in this study differ from previous literature that focused on analyzing solely forum posts~\cite{holt_subcultural_2007, benjamin_securing_2012, zhang_classification_2015, fang_exploring_2016}, actors' networks~\cite{decary-hetu_social_2012, samtani_using_2016, samtani_exploring_2017} or both \cite{abbasi_descriptive_2014, grisham_identifying_2017, marin_mining_2018 ,johnsen_identifying_2020, huang_hackerrank_2021}. This is because this study focuses on finding actors with technical expertise in attack patterns.

Still, the results of the study align with this previous literature, as key expert actors identified - the professionals - are active and senior actors~\cite{holt_techcrafters_2008, benjamin_securing_2012, abbasi_descriptive_2014,zhang_classification_2015,samtani_using_2016, grisham_identifying_2017,johnsen_identifying_2020, huang_hackerrank_2021} who are specialized in their discussions on attack patterns~\cite{abbasi_descriptive_2014, fang_exploring_2016} and have a high commitment percentage.

However, these actors form only 4\% of all actors studied; suggesting that they represent a very small fraction of the study population. Moreover, the study population represents a filtered sample with actors who posted about a CVE that had a related CAPEC at least four times (N=359). Hence, key expert actors represent a tiny part of the whole cybercrime population.  This finding aligns with previous literature on the key hacker identification problem stating that key actors make up only a small proportion of their platform's population~\cite{marin_community_2018}. 

The method developed in this study reduced the population of interest for intelligence production to just a small fraction of the final dataset. Hence, identifying a population of key threat actors comprising of a tiny proportion of the initial population can reshape resource allocation in cyber threat intelligence production, streamlining efforts for a more effective intelligence.

This is the main contribution of this study to the key hacker identification problem. We do not claim that the problem is solved, but we offer a way to detect key expert actors who need to be monitored. Further research should look into their influence in cybercrime forums as well as their success in conducting their criminal activities. 

\subsubsection{Highlighting the Large Presence of Amateurs with Regard to Their Technical Expertise}
Finally, the use of Bouchard and Nguyen's~\cite{bouchard_professionals_2011} framework -with the added activity rate variable- provides a better understanding of the study population with regards their technical expertise. It also suits well the relative approach on expertise~\cite{hoffman_eliciting_1995, hoffman_how_1998, chi_two_nodate, nee_review_2015} as it offers an overview of the distribution of technical expertise from novices [amateurs] to experts [professionals].

In this study, pro-amateurs account for 31.20\%. These actors have the potential to become experts as they already display high skills. They could additionally be monitored given their potential. This is especially true for those in cluster 1 with high skill, high commitment but low activity rate.  

Average career criminals account for 10\% of the sample and have low skill but high commitment. Since they have high commitment, they may have the will to become experts and sharpen their skills. Hence, they could be monitored as well, but to a lesser extent. 

Finally, amateurs are that those who score low on skill level and commitment account for 54.87\%. This means that more than half of the sample includes individuals who are far from experts: they are the novices or the rookies. Fewer resources can be put to their monitoring. This finding also aligns with previous research that shows the cybercrime industry is populated with lowly skilled individuals~\cite{collier2021cybercrime, paquet2022motivations}. 

\section{Limits and Further Research}
\label{sec:Limits}
Despite the strengths of this study, several limitations warrant consideration. First, this study uses CVEs as proxies for CAPECs, relying on MITRE’s CVE-CAPEC mapping and its accuracy. While this mapping lends credibility to the approach, potential information loss in the CVE-CAPEC association represents a limitation to this study. Next, this research uses MITRE’s skill level required metric for analysis, relying on it as the basis for our skill level calculations. However, since MITRE’s precise computation process for the skill level required metric isn’t publicly available; we rely on what could be considered a ‘black box’.

The imputation of skill level required values to valueless CAPECs also constitutes a limit of this study. Although the CAPEC framework’s hierarchical structure was used for imputation, both approaches, using the child or parent skill level, involve some level of estimation, which may not accurately reflect the true skill level of a CAPEC. Moreover, each CAPEC was then assigned its highest scenario’s value. This way, overestimating CAPECs skill level required, thus impacting actor’s skill level, is a possibility and a limit of this study. However, this approach prioritizes caution, as overestimating skill levels helps avoid underestimating potentially dangerous actors, ensuring they are captured in the analysis rather than slipping through undetected. Exploring alternative sources or methods for assessing required skill levels beyond MITRE’s metrics is a promising direction for future research.

This work proposed an objective assessment of a skill level in the sense that it avoids any human biases behind the qualitative analysis of actors’ content. Both metrics (skill level and commitment) have a threshold specifically suited to our sample's distribution, making them unsuitable for different samples. Future research could develop universally applicable skill and commitment metrics. 

The adaptation of the criminological framework has its own considerations. Without direct interviews to measure expertise, we relied on proxy variables, leading to a theoretical assessment of expertise based on recorded and available forum activities. The expertise measured is also static, representing a single point in time, and relies on forum presence. Future research could examine the evolution of expertise over time and investigate overlaps in key actors identified by different methods, comparing their positions in the broader cybercrime ecosystem.

This research acknowledges the possibility of identifying cybersecurity analysts, forum administrators, or law enforcement investigators as part of the Pro-Amateurs group. The aim of this study is to identify participants of interest from a cyber threat intelligence perspective, rather than to differentiate their specific roles. Distinguishing undercover investigators from malicious actors is beyond the scope of this study’s quantitative methods, as both may exhibit similar traits. As a result, a cybersecurity professional and a malicious actor might be categorized similarly in our study. Further qualitative analysis would be necessary to differentiate these roles, but this falls outside the scope of this research. Future research could dive deeper into the full content 
of those identified as Professionals and Pro-Amateurs to understand their role and study 
more closely this population identified as key.

Our framework differs from previous literature in the variables and metrics used to identify key actors. First, our data only had actors’ posts without tracking replies or the progression of discussion threads, limiting our ability to incorporate centrality measures.  Second, focusing solely on CVE mentions, without considering post content, sets this study apart from other key hacker identification research. This approach excludes comprehensive insights into an actor’s full forum activity, as posts are analyzed independently of their discussion threads, resulting in a loss of contextual information. Consequently, our metrics rely exclusively on specialized technical posts and may overlook valuable contributions without CVE mentions.

However, relying on CVE mentions for data collection ensures discussions are about direct vulnerabilities, reducing noise and trivial content. While it causes the loss of potential key posts and actors not referencing CVEs, it provides a focused dataset, enabling faster and more objective metric processing. This approach is more scalable and less prone to human biases. To account for the unspecialized content of actors, future research could collect actors’ non-specialized posts on top of those mentioning a CVE to study actors’ whole contribution in their community.

\section{Conclusion}
\label{sec:Conclusion}
This study identifies \textit{experts} in attack patterns \textit{across cybercrime forums}. Expertise in attack patterns is crucial for cybercrime intelligence, as it focuses on targeting actors with the knowledge and skills to attack enterprises.  Specifically, drawing on Bouchard and Nguyen~\cite{bouchard_professionals_2011}'s framework and previous literature, this study addresses the key hacker identification problem by, first, identifying communities interested in similar attack patterns across cybercrime forums and second, pinpointing their related key expert actors. 

To do so, we first identified areas of technical expertise in the form of communities of interest towards attack patterns. Leveraging CVE mentions from actors’ posts and their corresponding CAPECs, we built a bimodal network before using the Leiden algorithm to identify communities of interest. Then, using the K-means, we detected key actors based on their technical expertise. Resulting clusters were then interpreted in light of the Bouchard and Nguyen's~framework \cite{bouchard_professionals_2011} .

The results highlight that groups of actors interested in similar attack patterns exist and that actors with high skill and high commitment, the key expert actors, represent a small proportion, 4\% of the study population. They are the ones who should be monitored. Finally, the study highlight that a large proportion of the study population, more than half, displays little technical expertise, they are amateurs. Further research should focus on developing effective monitoring strategies for these key expert actors.

\section*{Acknowledgment}

This research was conducted as part of a MITACS research internship in collaboration with Secureworks, who assisted with data collection as well as with the analysis.

% \section*{References}

\bibliographystyle{plain}
\bibliography{recension}
\end{document}